\documentclass[12pt, twoside]{article}   
\usepackage[super,sort,comma]{natbib}
\usepackage{multirow}
\usepackage{amsmath}
\usepackage{amsfonts}
\usepackage{mathtools}
\usepackage{fancyhdr}		
\usepackage{enumitem}      

\DeclareMathOperator{\clip}{clip}
\DeclareMathOperator*{\argmin}{arg\,min}

\usepackage[section]{placeins}   %

\usepackage{graphicx}
\usepackage{subcaption}

\makeatletter \renewcommand\@biblabel[1]{$^{#1}$} \makeatother
\newcommand{\cen}[1]{\begin{center} #1 \end{center}}

\usepackage[mathlines]{lineno}

\usepackage{hyperref}
\hypersetup{ colorlinks,
    citecolor=blue,
    filecolor=blue,
    linkcolor=blue,
    urlcolor=blue
}

\usepackage{xcolor}

\definecolor{gray}{rgb}{0.6,0.6,0.6}
\definecolor{red}{rgb}{0.85,0,0}
\definecolor{green}{rgb}{0,0.85,0}
\definecolor{blue}{rgb}{0,0,0.85}
\definecolor{beige}{rgb}{0.92,0.87,0.78}
\usepackage[all]{hypcap}    

\begin{document}

\cen{\sf {\Large {Automating proton PBS treatment planning for head and neck cancers using policy gradient-based deep reinforcement learning} \\  
\vspace*{10mm}
Qingqing Wang$^1$, Chang Chang$^{1, 2}$} \\
$^1$Department of Radiation Medicine and Applied Sciences, University of California at San Diego, La Jolla, California, USA \\
$^2$California Protons Cancer Therapy Center, San Diego, California, USA
}

\pagenumbering{roman}
\setcounter{page}{1}
\pagestyle{plain}
Corresponding author: Chang Chang. Email: chang2@health.ucsd.edu\\
\begin{abstract}
\noindent {\bf Background:} 
Proton pencil beam scanning (PBS) treatment planning for head and neck (H\&N) cancers is a time-consuming and experience-demanding task where a large number of potentially conflicting planning objectives are involved. 
Deep reinforcement learning (DRL) has recently been introduced to the planning processes of intensity-modulated radiation therapy and brachytherapy for prostate, lung, and cervical cancers. 
However, existing DRL planning models are built upon the Q-learning framework and rely on weighted linear combinations of clinical metrics for reward calculation.
These approaches suffer from poor scalability and flexibility, 
i.e. they are only capable of adjusting a limited number of planning objectives in discrete action spaces and therefore fail to generalize to more complex planning problems.\\
{\bf Purpose:} 
Here we propose an automatic treatment planning model using the proximal policy optimization (PPO) algorithm in the policy gradient framework of DRL and a dose distribution-based reward function for proton PBS treatment planning of H\&N cancers. \\
{\bf Methods:} 
The planning process is formulated as an optimization problem.
A set of empirical rules is used to create auxiliary planning structures from target volumes and organs-at-risk (OARs), 
along with their associated planning objectives. 
Special attention is given to overlapping structures with potentially conflicting objectives.
These planning objectives are fed into an in-house optimization engine to generate the spot monitor unit (MU) values.
A decision-making policy network trained using PPO is developed to iteratively adjust the involved planning objective parameters.
The policy network predicts actions in a continuous action space and guides the treatment planning system to refine the PBS treatment plans using a novel dose distribution-based reward function. 
A total of 34 H\&N patients (30 for training and 4 for test) and 26 liver patients (20 for training, 6 for test) are included in this study to train and verify the effectiveness and generalizability of the proposed method. \\
{\bf Results:} 
Proton H\&N treatment plans generated by the model show improved OAR sparing with equal or superior target coverage when compared with human-generated plans. 
Moreover, additional experiments on liver cancer demonstrate that the proposed method can be successfully generalized to other treatment sites. \\
{\bf Conclusions:} 
The automatic treatment planning model can generate complex H\&N plans with quality comparable or superior to those produced by experienced human planners.
Compared with existing works, our method is capable of handling more planning objectives in continuous action spaces. 
To the best of our knowledge, this is the first DRL-based automatic treatment planning model capable of achieving human-level performance for H\&N cancers. \\
\end{abstract}
\setlength{\baselineskip}{0.7cm}      
\pagenumbering{arabic}
\setcounter{page}{1}
\pagestyle{fancy}
\section{INTRODUCTION}

Treatment planning is a patient-specific optimization problem solved by human planners using treatment planning systems (TPS).
Quality of the resultant plans is determined by the experience of the planners, algorithms in the TPS, and time available for planning.
In the planning process, human planners first create numerous auxiliary planning structures and their corresponding objectives according to the clinical requirements. 
Using these objectives as inputs, the optimization engine in TPS calculates the machine deliverable parameters, such as the dwell time for brachytherapy and spot MU values for proton PBS. Dose-volume histograms (DVHs) are computed from the resultant dose distributions to help evaluate the quality of generated plans. 
Typically, objectives in TPS involve a set of hyper-parameters, such as objective weights, dose limits and volume constraints. 
If generated plans do not satisfy the clinical criteria, human planners have to modify planning structures and their objectives before conducting the optimization again. 
The interaction is repeated until a plan satisfying all clinical constraints is produced. 
This iterative process is time-consuming and labor-intensive, and it may take an experienced planner hours for complex treatment sites such as H\&N.
Therefore, an automatic treatment planning system capable of generating high-quality plans without the need for human interference is keenly desired. 
However, this expertise-demanding decision-making task is challenging and despite extensive efforts over the last few decades its resolution remains illusive. 

Existing automatic treatment planning techniques can be categorized into conventional machine learning approaches and more advanced deep learning-based methods.
In the first group, recursive random search~\cite{A04}, decision-function~\cite{A05}, fuzzy inference~\cite{A06}, regression model~\cite{A07} and so forth were employed to determine a small group of planning objectives in an outer parameter adjustment loop, which was built on top of an inner iteration used to solve the optimization problem. 
Recently, deep learning was introduced to automate treatment planning and it has undoubtedly advanced the development of automatic TPS. 
For the task of adjusting planning objective parameters, 
deep reinforcement learning (DRL), the most sophisticated decision-making method, was highly favored~\cite{A08,A09,A10,A11,A16,A18,A19}. For examples, Shen et al~\cite{A08, A09} provided a proof-of-principle study on learning the weight-tuning intelligence of a deep policy network trained in the trial-and-error manner of Q-learning~\cite{A13}. 
However, the size of their policy network increased linearly with the number of planning objective parameters, 
resulting in limited scalability. 
Subsequent study from the same group proposed another hierarchical policy network~\cite{A10} which adjusted one parameter for one structure at a time, 
therefore improving the model's scalability at the expense of efficiency. 
Pu et al~\cite{A11} built a dwell time prediction model for cervical brachytherapy by combining two variants of deep Q-learning. This approach was able to solve complex optimization problems, but it used a fixed time adjustment step of 0.1 seconds, instead of employing adaptively varied adjustment steps.  
Similarly, Hrinivich et al~\cite{A18} employed deep Q-learning in their volumetric modulated arc therapy (VMAT) machine parameter optimization, and Zhang et al~\cite{A19} demonstrated an interpretable and reproducible treatment planning bot for pancreas SBRT. 

Although encouraging achievements have been accomplished from the above explorations, existing DRL-based automatic treatment planning still suffers from the following limitations: 
1) \textbf{scalability}: existing methods are all developed using Q-learning, whose network size grows linearly with the number of parameters to be adjusted. 
Therefore, they are only suitable for simpler treatment sites involving a small number of planning objectives; 
2) \textbf{flexibility}: existing methods predict actions in a discrete space, which limits their policy networks to a finite number of actions at fixed steps, seriously affecting adjustment efficiency and precision. 
As pointed out in earlier works~\cite{A08, A11, A16}, using continuous action space is a potential solution to this problem; 
3) \textbf{generalizability}: existing methods rely heavily on weighted combinations of clinical metrics when calculating rewards for predicted actions, 
failing to make reasonable trade-offs when a diverse number of target volumes, OARs and prescription levels are involved across different treatment sites. 
As a result, existing approaches struggle to generalize to complex sites such as H\&N. 

To address these challenges, in this pilot study we propose an automatic treatment planning method using Proximal Policy Optimization (PPO)~\cite{A22}, which predicts actions in a continuous action space. 
First, a group of empirical rules are carefully designed to create auxiliary planning structures and their associated non-conflicting objectives for an in-house optimization engine, 
which generates the first iteration of spot MU values for the plan.  
Secondly, a Transformer-based~\cite{A24} PPO policy network is used to adjust the planning objective parameters of a set of dynamically selected OARs. 
This policy network is trained by interacting with our treatment planning environment, which utilizes a novel dose distribution-based reward function to guide the policy network in navigating the trade-offs amongst dose distributions of the target volumes and the OARs. 
The proposed model automatically generates patient treatment plans with human-level performance for H\&N  cancers. 
Generalization of this model to other treatment sites such as liver is also demonstrated.  

\section{MATERIALS AND METHODS}
\subsection{Patient cohorts}

Patient cohorts are selected from those treated at California Protons Cancer Therapy Center from 2017 to 2023. 
The dataset consists of 34 bilateral and ipsilateral H\&N cancer patients (30 for training and 4 for test), 
and 26 liver cancer patients (20 for training and 6 for test). 
Patients are treated with either single-field optimized (SFO) or multi-field optimized (MFO)~\cite{Frank2014} plans that were manually planned by experienced dosimetrists. 
The number of treatment fractions ranges from 5 to 60~{\it bid}. 
Up to 5 treatment beams per plan are used and field-specific targets are created to constrain spot placements~\cite{A21}. 
The model is designed to accommodate a maximum of 4 different prescription dose levels found in the H\&N cohort.  
Prescription doses, beam angles, field targets and target volumes in our experiments are kept identical to those used in clinical plans. 
Spots with MU values less than the allowed minimum MU per spot are removed from the generated plans. 
The deliverable minimum MU per spot is a characteristic of the proton hardware and it is set to 3MU in our clinic.

\subsection{Automatic treatment planning}
\begin{figure}[h]
   \begin{center}
   \includegraphics[width=0.85\textwidth]{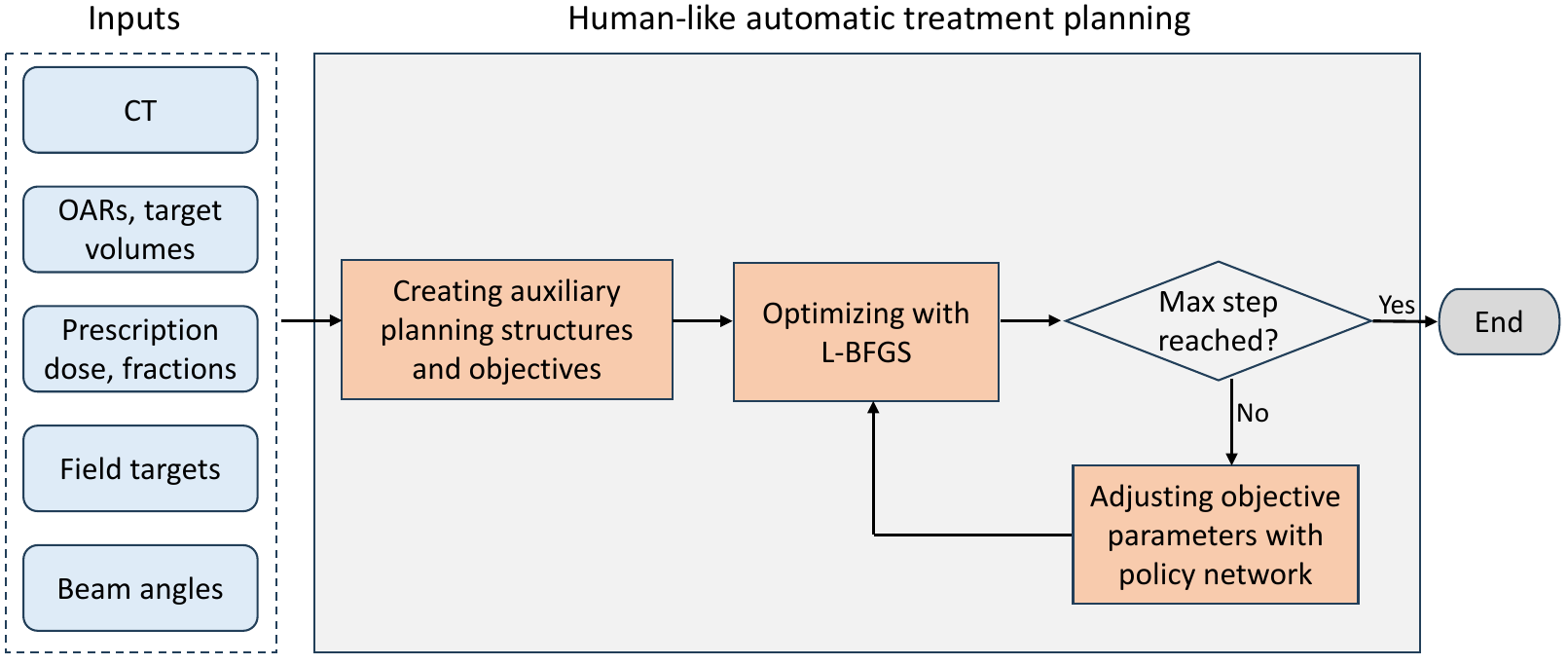}
   \caption{Framework of proposed automatic treatment planning method.
   \label{fig:framework} 
    }  
    \end{center}
\end{figure}
 
Figure~\ref{fig:framework} depicts the design of the proposed automatic treatment planning model which is based on the typical treatment planning workflow. 
Auxiliary planning structures and associated objectives are first created with a set of empirical rules. 
Machine deliverable spot MU values are initialized by an in-house optimization engine using the L-BFGS algorithm. 
Planning objective parameters are then iteratively adjusted by the policy network to reach the final spot MU values. 
To build a standard pipeline suitable for all patients, all OARs found in this patient cohort are included in the model. 
Objectives of target volumes are fixed with large weights and stringent dose limits so that target coverages are prioritized for all patients at initialization. 
The goal of our policy network is to gradually reduce OAR doses with minimal impact on target coverages and this is achieved by deliberately adjusting planning objective parameters in continuous action spaces.  
Here, the policy network is trained with the advanced PPO framework under the guidance of a novel dose distribution-based reward function.

\subsection{Creation of auxiliary planning structures and objectives} 
\label{sec:EmpiricalRules}
\begin{figure}[h]
   \begin{center}
   \includegraphics[width=0.6\textwidth]{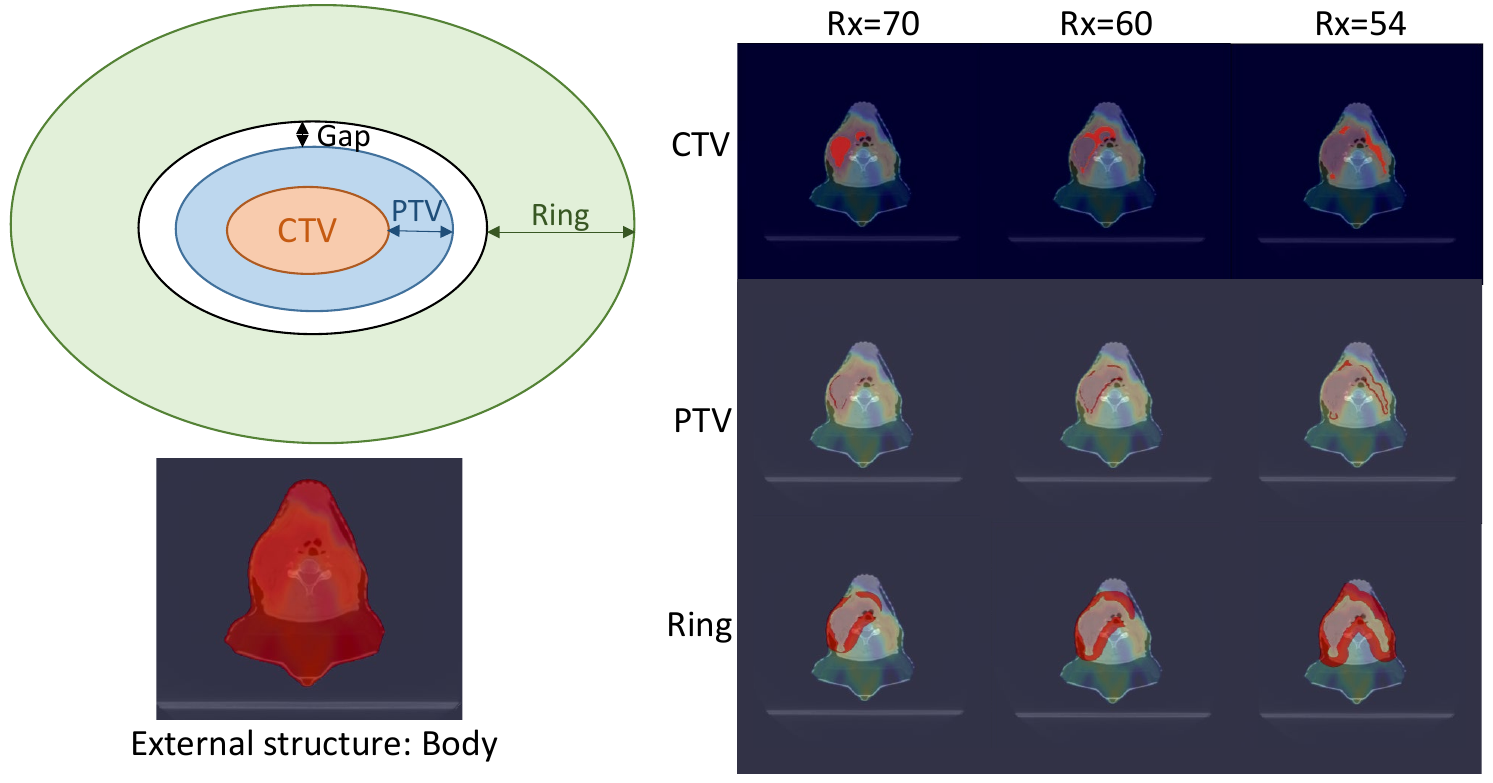}
   \caption{Auxiliary planning structures (masked in red) for target volumes.
   \label{fig:controlStructures} 
    }  
    \end{center}
\end{figure} 

For target volumes, we create a 3mm expansion from the clinical target volume (CTV) as PTV, a 2mm expansion from PTV as Gap, and a 20mm expansion from Gap as Ring (see Figure~\ref{fig:controlStructures}).
In addition, 32 OARs are accounted for in the current H\&N model.
Following established practice, critical OARs such as  ``Spinal Cord'' and ``Brainstem'' are given higher priority than target volumes. 
To summarize, all target and OAR structures are classified into three groups: 

\begin{description}[wide=0\parindent,noitemsep,nolistsep]
  \item[Level 1:] ``Brainstem'', ``Brainstem Core'', ``Spinal Cord'', ``Chiasm'', ``Optic Nerves'';
  \item[Level 2:] CTV, PTV;
  \item[Level 3:] Ring, ``Bone Mandible'', ``Brachial Plexus'', ``Brain'', ``Cavity Oral'', ``Cochlea'', ``Ear IntMid'', ``Esophagus'', ``Gland Submandibular'', `Gland Thyroid'', ``Larynx'', ``Lips'', ``Lungs'', ``Muscle Constrict'', ``Parotid'', ``Spinal Canal'', ``Skin'', ``Eye'',``Oropharynx'', ``Gland Lacrimal'', `Lens'', ``Pituitary'', ``Lobe Temporal'', ``Cornea'', ``Retina'', ``Vocal Cords'', ``Nasopharynx'', ``Carotid''
\end{description}

In general, we create objectives for the optimization engine by setting dose limits on the maximal dose ($D_{max}$), minimal dose ($D_{min}$) and mean dose ($D_{mean}$) of the associated structures. Specifically, target volumes require dose limits for both $D_{max}$ and $D_{min}$ so that their doses are constrained within a proper range. 
Rings and OARs require dose limits for $D_{max}$ or $D_{mean}$ so that normal tissues are appropriately spared. 
Apparently, there will be conflicts when CTVs and PTVs of different prescription levels overlap, 
and when CTVs and PTVs intersect with Rings and OARs. 
Conflicts between objectives pose challenges to the optimization and usually lead to undesirable results. 
Therefore, designing non-conflicting objectives for the optimization engine is vital, especially when multiple prescription dose levels are involved. 
To address this issue, we define the following empirical rules for planning structures and associated objectives:  
 \begin{enumerate}
	\small
	\item For all structures, portions outside the ``external'' or ``Body'' contour will be removed;
	\item If CTV (or PTV) intersects with ``Skin'', the overlap should be subtracted from CTV (or PTV);
	\item If CTV$_i$ intersects with CTV$_j$ and $\mathsf{Rx}_i > \mathsf{Rx}_j$ ($\mathsf{Rx}$ denotes prescription dose), the overlap should be subtracted from CTV$_j$;
	\item If PTV$_i$ intersects with CTV$_j$, the overlap should be subtracted from PTV$_i$ as CTVs are always of higher priority than PTVs;
	\item If PTV$_i$ intersects with PTV$_j$ and $\mathsf{Rx}_i > \mathsf{Rx}_j$, the overlap should be subtracted from PTV$_j$;
	\item If Ring intersects with CTV (or PTV), the overlap should be subtracted from Ring;
	\item If Ring$_i$ intersects with Ring$_j$ (or any OARs), the overlap belongs to both Ring$_i$ and Ring$_j$ (or OARs) since there is no conflict between their objectives;
	\item When CTV$_i$ (or PTV$_i$) intersects with OAR$_j$, if the priority of OAR$_j$ is higher, the overlap should be subtracted from CTV$_i$ (or PTV$_i$); otherwise, the overlap should be subtracted from OAR$_j$;
	\item $D_{max}$ objectives are used for structures related to Ring, ``Bone Mandible'', ``Brachial Plexus'', ``Brain'', ``Cochlea'', ``Spinal Canal'', ``Skin'', ``Eye'', ``Pituitary'', ``Lobe Temporal'', ``Retina'', ``Vocal Cords'', ``Carotid'', ``Brainstem'', ``Brainstem Core'', ``Spinal Cord'', ``Optic Chiasm'' and ``Optic Nerve'', and $D_{mean}$ objectives are used for the rest of the OARs;
	\item Dose limits and weights for the $D_{min}$ and $D_{max}$ objectives of CTVs are not adjustable and they are fixed at $1.02 \cdot \mathsf{Rx}$, 
	$1.03 \cdot \mathsf{Rx_{max}}$, 2.0 and 2.0, respectively. 
	Likewise, these parameters are fixed at $0.97 \cdot \mathsf{Rx}$, $1.02 \cdot \mathsf{Rx_{max}}$, 1.0 and 1.0 for the $D_{min}$ and $D_{max}$ objectives of PTVs, respectively;
	and $0.93 \cdot \mathsf{Rx}$ and 1.0 for the $D_{max}$ objectives of Rings. 
	$\mathsf{Rx_{max}}$ represents the maximal prescription dose amongst all target volumes;
	\item Dose limits for OARs' objectives are initialized to relatively larger values in order to minimize their interference with target coverages in the first iteration of the optimization. Specifically, a UNETR-like~\cite{A29} dose prediction model is trained in-house to estimate the dose distribution for each patient and the initial dose limits for each OAR are randomly sampled to be 10\% to 30\% larger than the UNETR’s estimation. Weights for the OARs' objectives are initialized to 1.0 and will be iteratively adjusted by our policy network, together with their associated dose limits.
 \end{enumerate}

Overall, the model is designed to accommodate up to 4 prescription dose levels, 50 planning structures, 58 planning objectives and 76 adjustable planning objective parameters. 
By contrast, the maximal number of objectives and adjustable parameters in existing works are 16 and 48, respectively, with only one prescription dose level allowed~\cite{A08, A09, A10, A11, A16, A19}.   

\subsection{Plan optimization with L-BFGS}
\label{opt_engine}
Radiation therapy treatment planning can be formulated as an optimization problem:
\begin{align}
\widehat{X} 
&=\argmin_{X}\,L(X) \notag \\ 
\begin{split}
&=\argmin_{X} \Biggl[ \sum_{\;i\in \mathcal{S}_{D_{max}}} \!\!\!w^i \biggl(\frac{1}{N^i}\sum\bigl\lVert M^iX - D_{max}^i \bigr\rVert^2_+ \biggr) \\
&\quad + \sum_{j \in \mathcal{S}_{D_{min}}} \!\!\!w^j \biggl( \frac{1}{N^j} \sum\bigl\lVert M^j X - D_{min}^j \bigr\rVert^2_- \biggr)  
    + \sum_{k\in \mathcal{S}_{D_{mean}}} \!\!\!w^k\biggl( \Bigl\lVert\frac{1}{N^k}\sum M^k X - D_{mean}^k \Bigr\rVert^2_+ \biggr) \Biggr] ,
\label{eq:eq1}
\end{split}
\end{align}
where $L(X)$ is the loss function given the argument $X$ to be optimized, 
and for proton PBS plans, $X$ is the spot MU values.
$\mathcal{S}_{D_{max}}$, $\mathcal{S}_{D_{min}}$ and $\mathcal{S}_{D_{mean}}$ represent sets of planning structures with objectives on maximum dose, minimum dose, and mean dose, respectively. 
The corresponding weights for these objectives are denoted as $w^i$, $w^j$ and $w^k$, and the corresponding dose limits are denoted as $D_{max}^i$, $D_{min}^j$ and $D_{mean}^k$, respectively.
The number of voxels in each planning structures $i$, $j$, $k$ in the sets $\mathcal{S}_{D_{max}}$, $\mathcal{S}_{D_{min}}$ and $\mathcal{S}_{D_{mean}}$ are represented as $N^i$, $N^j$ and $N^k$, and their corresponding dose influence matrices~\cite{Unkelbach2016, Ma2018} are denoted as $M^i$, $M^j$ and $M^k$, respectively. 
Moreover, $\lVert\cdot\rVert^2_+$ and $\Vert\cdot\rVert^2_-$ are standard $l_2$-norms that compute only the positive and the negative elements, respectively.   

Here we use the well-known quasi-Newton method Limited-memory Broyden–Fletcher–Goldfarb–Shanno (L-BFGS)~\cite{A23} to solve the optimization problem. 
L-BFGS is specifically designed to solve problems involving large numbers of variables, 
where computing and storing the whole Hessian matrix are not practical. 
In Newton optimization methods like L-BFGS, $X$ can be iteratively optimized with
\begin{equation}
 X_{t+1} = X_t - \psi H_t \, g_t \,,
\label{eq:eq2}
\end{equation}
where $\psi$, $g_t$ and $H_t$ are the step length, the gradient and the inverse Hessian at point $X_t$, respectively. Since the inverse Hessian is generally dense, it is expensive to compute and store the inverse Hessian $H_t$ when $X$ is of high dimensionality. 
Therefore, L-BFGS approximates the inverse Hessian matrix at every iteration by
\begin{equation}
\begin{split}
H_{t+1} &= \left( V^T_t \cdots V^T_{t-m+1} \right) H^0_{t+1}\left( V_{t-m+1} \cdots V_t \right) \\
&\quad + \rho_{t-m+1} \left( V^T_t \cdots V^T_{t-m+2} \right) x_{t-m+1}x^T_{t-m+1} \left( V_{t-m+2} \cdots V_t \right) \\
					 &\quad + \rho_{t-m+2} \left( V^T_t \cdots V^T_{t-m+3} \right) x_{t-m+2}x^T_{t-m+2} \left( V_{t-m+3} \cdots V_t \right) + \cdots + \rho_t x_t x^T_t \,,
\end{split}
\label{eq:eq4}
\end{equation}
where $V_t = I - \rho_t y_t x^T_t$, $\rho_t = \frac{1}{y^T_t x_t}$, $x_t = X_{t+1} - X_t$ and $y_t = g_{t+1} - g_t$. Here, $H^0_{t+1}$ is the initial Hessian approximation at point $X_{t+1}$, and in practice it can be set to $\frac{x^T_t y_t}{y^T_t y_t} I$, where $I$ is the identity matrix.
As seen in Eq.~\ref{eq:eq4}, 
L-BFGS alleviates the memory requirement by using a limited number, i.e. $m$, 
of the vector pairs $\{x_i, y_i\}$, $i=t,...,t-m+1$ for $H_{t+1}$. In our experiments, $m$ is set to 10.

\subsection{Planning objective parameter adjustment with DRL}
In a typical planning pipeline, human planners manually adjust planning objective parameters, 
i.e. $w^i$, $w^j$, $w^k$ and $D^i_{max}, D^j_{min}, D^k_{mean}$ in Eq.~\ref{eq:eq1}, to improve plan quality iteratively. 
We automate this time-consuming and experience-demanding process with the advanced reinforcement learning algorithm, i.e., PPO~\cite{A22}.

\textbf{Algorithm:}
PPO is from the well-known policy gradient family of algorithms, 
which alternates between sampling data through interactions with a task-specific environment and training a policy using stochastic gradient ascent. 
The basic idea is to optimize the policy by computing the gradients of expected rewards with respect to policy parameters. 
Assuming that the PPO agent receives a state $s_t$ at time step $t$, 
it then generates an action $a_t$ according to its current policy $\pi_\theta$. 
Here, $\pi$ is a policy mapping from states to actions and we use $\pi_\theta$ to denote a policy with parameters $\theta$. 
In response to $a_t$, the environment reaches a new state $s_{t+1}$ after applying $a_t$ to $s_t$ and returns a scalar reward $r_t$ as the feedback for taking action $a_t$. 
This process continues until a terminal state is reached at time step $T$.
Here we denote the policy used for data sampling as $\pi_{\theta'}$, and the policy to be updated as $\pi_{\theta}$.
In each episode, $M$ trajectories $\tau^i=\{a_1^i, a_2^i, \cdots, a_T^i\}$, $i=1\cdots M$, are sampled with the policy $\pi_{\theta'}$, 
and PPO updates $\theta$ multiple times using the gradient estimator
\begin{equation}
	\hat{g}_{_{\mathrm{PPO}}}^{\theta'}(\theta) = \frac{1}{M}\sum_{i=1}^M\sum_{t=1}^T \frac{\pi_\theta\left(a_t^i|s_t^i\right)}{\pi_{\theta'}\left(a_t^i|s_t^i\right)} A^{\pi_{\theta'}}\bigl(s_t^i, a_t^i\bigr) \nabla \log \pi_\theta\bigl(a_t^i|s_t^i\bigr) 
\label{eq:eq10}
\end{equation}
where
\begin{align*}
A^{\pi_{\theta'}}\bigl(s_t^i, a_t^i\bigr) &= r^i_t + V^{\pi_{\theta'}}\bigl(s^i_{t+1}\bigr) - V^{\pi_{\theta'}}\bigl(s^i_t\bigr), \\
V^{\pi_{\theta'}}\bigl(s^i_{t}\bigr) &= \mathbb{E}_{s^i_{t+1:\infty},a^i_{t:\infty}}\Biggl[\sum_{l=0:\infty}{r^i_{t+l}}\Biggr].
\end{align*}

Here, the advantage function $A^{\pi_{\theta'}}\bigl(s_t^i, a_t^i\bigr)$ represents the advantage of action $a_t^i$ at state $s^i_t$ in the $i$-th trajectory, compared to the default behavior of policy $\pi_{\theta'}$. 
The value function $V^{\pi_{\theta'}}\bigl(s^i_t\bigr)$ is the expected accumulative reward of state $s_t^i$ under policy $\pi_{\theta'}$, 
representing the performance of this policy. 
The policy function $\pi_\theta\left(a_t^i|s_t^i\right)$ represents the probability of policy $\pi_\theta$ choosing action $a_t^i$ at state $s_t^i$.
The importance sampling ratio $\frac{\pi_\theta\left(a_t^i|s_t^i\right)}{\pi_{\theta'}\left(a_t^i|s_t^i\right)}$ is introduced to allow multiple policy updates per sampling rollout collection, thereby making training more data-efficient.
Intuitively, when $a_t^i$ is a better-than-average action, i.e., $ A^{\pi_{\theta'}}\bigl(s_t^i, a_t^i\bigr) > 0$, the gradient term $A^{\pi_{\theta'}}\bigl(s_t^i, a_t^i\bigr) \nabla \log \pi_\theta(a_t^i|s_t^i)$ points in the direction of increasing the probability of yielding action $a_t^i$. Similarly, if $a_t^i$ is a worse-than-average action, i.e., $A^{\pi_{\theta'}}\bigl(s_t^i, a_t^i\bigr) < 0$, the probability of yielding action $a_t^i$ will be decreased.
Using $\nabla f(x) = f(x)\nabla \log f(x)$, 
the objective function of PPO corresponding to the gradient estimator in Eq.~\ref{eq:eq10} is given by
\begin{equation}
J^{\theta'}_{\mathrm{PPO}}(\theta) = \frac{1}{M}\sum_{i=1}^M\sum_{t=1}^T \frac{\pi_\theta\left(a_t^i|s_t^i\right)}{\pi_{\theta'}\left(a_t^i|s_t^i\right)} A^{\pi_{\theta'}}\bigl(s_t^i, a_t^i\bigr) 
\label{eq:eq11}
\end{equation}
and $\theta$ can be optimized by maximizing $J^{\theta'}_{\mathrm{PPO}}(\theta)$ with gradient ascent.

\begin{figure}[h]
   \begin{center}
   \includegraphics[width=1.0\textwidth]{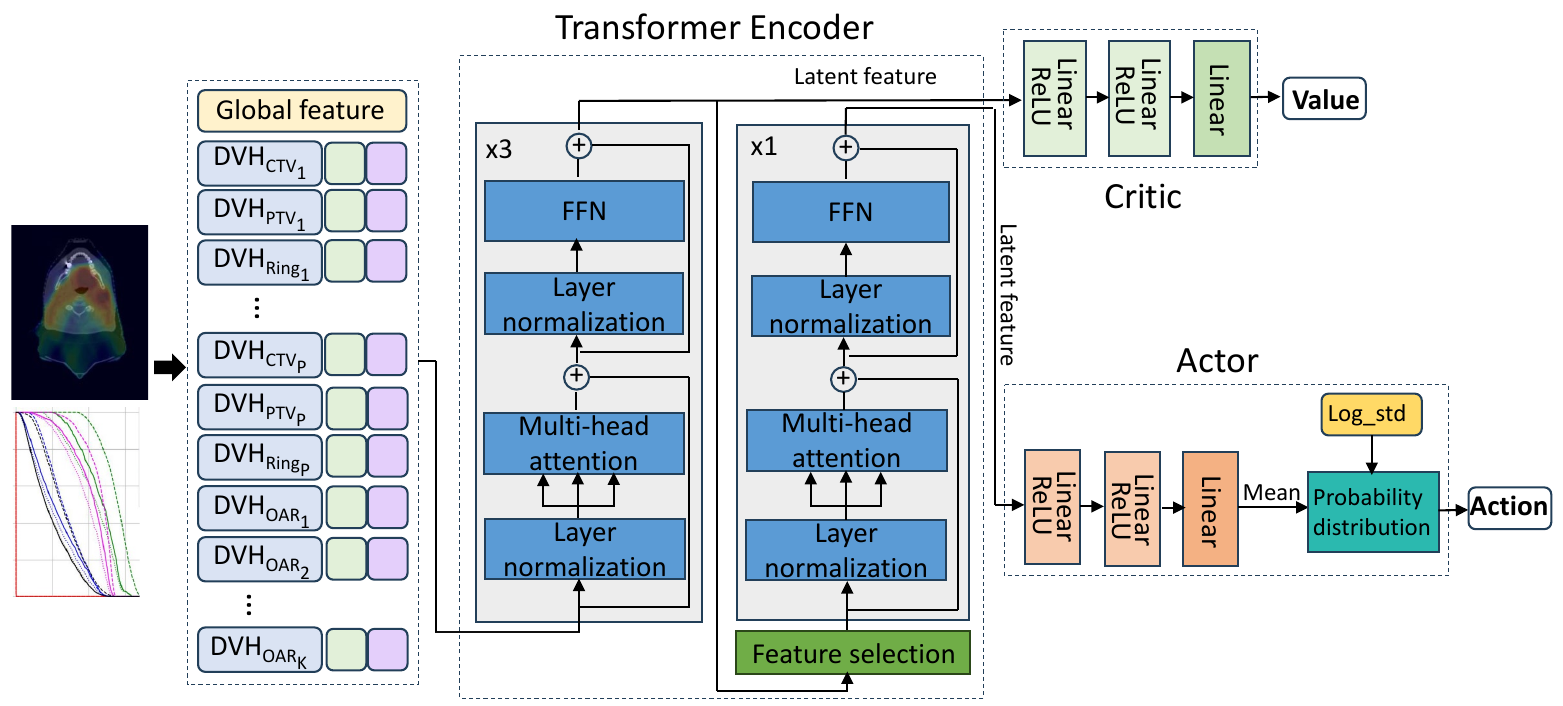}
   \caption{Transformer-based actor-critic agent. The actor predicts a distribution for action sampling and the critic evaluates actor's performance. ``Log\_std'' is the randomly initialized standard deviation of predicted distribution, and it is learned during the training procedure, together with other network parameters.    
   \label{fig:network} 
    }  
    \end{center}
\end{figure} 
\textbf{Actor-critic agent:} PPO is built upon the actor-critic architecture, 
where an actor plays the role of policy network $\pi$, and a critic estimates value $V^\pi(s_t)$ for state $s_t$ when calculating advantage $A^\pi(s_t,a_t)$. 
Inspired by the great potential of Transformers~\cite{A24}, we design a Transformer-based actor-critic agent. 
As illustrated in Figure~\ref{fig:network}, 
this agent consists of three components: a Transformer encoder that takes states as inputs and produces latent features for subsequent modules, 
an actor sub-network that generates a probability distribution for sampling actions, 
and a critic sub-network used to evaluate the performance of the actor. 
In our treatment planning environment, states are DVH features, which are obtained by sampling $N$ points uniformly along the dose axis of DVH curves. 
Corresponding percent volumes are taken as $N$-dimensional DVH features.
Planning structures' types, importance levels, scores (defined in Section~\ref{sec:reward}) and prescription doses are also considered as part of the states to provide additional information. Therefore, assuming $M$ planning structures are accounted for in each plan, the states have a dimensionality of $\left[M, N+4\right]$. 
Moreover, following the standard Transformer architecture, 
we also randomly initialize a learnable parameter with shape $\left[M, N+4\right]$ as positional encoding and another learnable parameter with shape $\left[1, N+4\right]$ as a global feature. 
The positional encoding is added to the states, and the global feature is attached on the top of the states to gather information from a global perspective. 
The resultant states and global features are encoded by the Transformer encoder, and the encoded global feature is then forwarded to the actor sub-network and the critic sub-network as latent features for further prediction.

The Transformer encoder is composed of multiple Transformer layers. 
Each layer consists of a multi-head self-attention module and a fully connected feed-forward network (FFN), around which residual connection and layer normalization are embedded. 
Assuming that there are $K$ adjustable planning objective parameters, 
the actor sub-network is supposed to predict a $K$-dimensional Gaussian distribution with a diagonal covariance matrix for action sampling. 
Specifically, it predicts the mean of the distribution from the input states. 
The standard deviation of the distribution, namely ``Log\_std'' in Figure~\ref{fig:network}, is randomly initialized and learned alongside other network parameters during training.  

In practice, H\&N proton PBS treatment planning involves a large number of adjustable parameters in planning objectives. 
The action space is therefore of high dimensionality and a large amount of data must be sampled from the environment in order to fully explore it. As a consequence, network training is computationally expensive. In addition, effective guidance is required to find positive samples in such a large exploration space, posing challenges to maintaining data balance. This ``curse of dimensionality'' is the main obstacle in applying reinforcement learning to many real-world problems.  
To address this issue, 
we reduce the dimensionality of the action space by adjusting only the planning objective parameters from a subset of planning structures that are dynamically selected by a feature selection module (see Figure~\ref{fig:network}).
For simplicity, this feature selection module directly selects planning structures with the lowest scores. 
Moreover, we also constrain the search space of actions to a limited range of $0 \sim 0.5$ and $-0.5 \sim 0$ for weight factors and relative dose limits, respectively. Given the action $a_t$ sampled from the predicted distribution at time step $t$, 
the adjustable planning objective parameters will be updated with $p_{new} = p_{old} * \left(1 + a_t\right)$, 
where $p_{old}$ and $p_{new}$ are planning objective parameters before and after adjustment.

\textbf{Network training:} To train the above network, we design the loss function $L_{_{\text{PPO}}}$ as
\begin{align}
L_{_{\text{PPO}}} &= L_{policy} + \alpha L_{value} + \beta L_{entropy} \notag\\
\begin{split}
&= \sum_{t=1}^T\Biggl[-\min\biggl(A^{\pi_{\theta'}}\left(s_t, a_t\right) \frac{\pi_\theta\left(a_t|s_t\right)}{\pi_{\theta'}\left(a_t|s_t\right)}, A^{\pi_{\theta'}}\left(s_t, a_t\right) \clip\Bigl(\frac{\pi_\theta\left(a_t|s_t\right)}{\pi_{\theta'}\left(a_t|s_t\right)}, 1-\mu, 1+\mu \Bigr)\biggr) \\
&\qquad\qquad  + \alpha \text{MSE}\Bigl(A^{\pi_{\theta'}}\left(s_t, a_t\right)+V^{\pi_{\theta'}}\left(s_t\right), V^{\pi_{\theta}}\left(s_t\right)\Bigr) - \beta\, \mathcal{H}\Bigl(\pi_{\theta}\left(\cdot|s_t\right)\Bigr)\Biggr] 
\end{split}
\label{eq:eq13}
\end{align} 
where $L_{policy}$ is the policy loss designed based on Eq.~\ref{eq:eq11}, $L_{value}$ is the Mean Square Error (MSE) loss used for updating the critic, and $L_{entropy}$ is the entropy loss used to enable stability and exploration.
Here, the operation $\clip\bigl( f(\cdot), 1-\mu, 1+\mu \bigr)$ limits the output of the function $f(\cdot)$ to the interval 
$[1-\mu, 1+\mu]$, in order to stabilize the training process by removing the incentive for larger differences between the old policy $\pi_{\theta'}$ and the new policy $\pi_\theta$.
The $\mathcal{H}\bigl(\pi_{\theta}(\cdot|s_t)\bigr)$ is the entropy of the current policy $\pi_{\theta}$ at state $s_t$. The idea of entropy regularization is derived from Soft Actor-Critic (SAC)~\cite{A33}, where the policy is incentivized to explore more widely, abandon unpromising avenues, and capture multiple modes of near-optimial behavior by maximizing both the expected returns and the expected entropy. Experiments conducted on SAC also showed that learning a stochastic policy with entropy regularization can significantly stabilize the training process. 
Moreover, we use Generalized Advantage Estimation (GAE)~\cite{A26} for advantage $A^{\pi_{\theta'}}\bigl(s_t, a_t\bigr)$, i.e.  
\begin{equation}
A^{\pi_{\theta'}}\bigl(s_t, a_t\bigr) \coloneq \widehat{A}_t^{\text{GAE}\left(\gamma,\lambda\right)} = \sum_{l=0}^\infty \left(\gamma\lambda\right)^l \Bigl(r_{t+l}+\gamma V^{\pi_{\theta'}}(s_{t+l+1}) - V^{\pi_{\theta'}}(s_{t+l})\Bigr) \,,
\label{eq:eq13_1}
\end{equation} 
where $\gamma \in [0, 1]$ is the discount rate and $\lambda \in [0, 1]$ is the trace-decay parameter used to control the compromise between bias and variance of the estimation. 
During training, planning objectives with weights and dose limits, 
either defined initially by the empirical rules or adjusted later by the policy network, are used for L-BFGS optimization. 
Spot MU values resulted from the optimization are used to compute the reward and the new state.
Based on the new state the policy network predicts the new action, which is then used to adjust the weights and the dose limits for the next iteration of L-BFGS optimization.
This process iterates $T$ times to generate a sequence of data $\bigl\{s_t, a_t, r_t, s_{t+1}\bigr\}$, with which the network is updated. 
The updated network is subsequently used to sample new sets of data sequences for training in the next episode.    

\subsection{Dose distribution-based reward function}
\label{sec:reward}

Treatment planning is essentially a process of balancing competing priorities between dose-volume requirements of the target volumes and the OARs. 
Prior DRL-based treatment planning studies simply employ a weighted sum of clinical metrics to score plans and calculate rewards. 
While straightforward, this strategy could lead to unreasonable trade-offs. 
For example, linearly combined $d_{2cc}$ values from OARs are used to calculate rewards $r_t = \sum_{i}w_i\bigl(d^i_{2cc}(s_{t+1}) - d^i_{2cc}(s_t)\bigr)$ in an earlier study~\cite{A08}.
This design resulted in identical rewards for a decrease in one OAR's $d_{2cc}$ from 60Gy to 50Gy and from 15Gy to 5Gy. 
However, in practice, clinicians are more likely to favor a 10Gy drop from 60Gy than from 15Gy, {\it ceteris paribus}. 
Indeed, a 10Gy reduction should be encouraged with a higher reward when, say, the rectum's $d_{2cc}$ is at 60Gy rather than at 15Gy due to their difference in toxicity risks. 
The irrationality of equal rewards for unequal clinical benefits becomes more pronounced with more OARs and target volumes involved. 
Therefore, a reasonable reward design that allows proper trade-offs between target volumes and OARs to be considered is critical for the DRL model.

\begin{figure}[h]
   \begin{center}
   \includegraphics[width=0.75\textwidth]{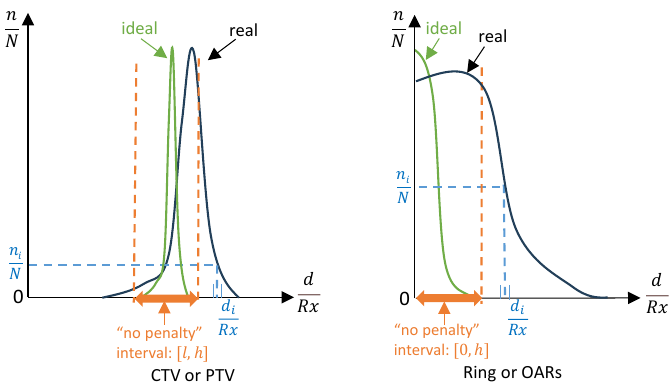}
   \caption{Dose distribution-based reward.    
   \label{fig:reward} 
    }  
    \end{center}
\end{figure}
Here we propose a novel reward function from the perspective of global dose distributions. 
As shown in Figure~\ref{fig:reward}, 
we set a ``no penalty'' interval $[l, h]$ relative to the prescription dose $\mathsf{Rx}$ for each planning structure, where all voxels within that structure receiving doses in the range of $\bigl[l\cdot\mathsf{Rx}, h\cdot\mathsf{Rx}\bigr]$ won't received a penalty. 
Penalty is only incurred for any voxels receiving doses outside that range. 
In our experiments, the ``no penalty'' intervals for CTVs and PTVs are set to $\bigl[1.0, 1.05\cdot\mathsf{Rx_{max}}/\mathsf{Rx}\bigr]$ and $\Bigl[0.95, 1.03\cdot\mathsf{Rx_{max}}/\mathsf{Rx}\Bigr]$, respectively.
For plans with multiple levels of prescription doses, each target volume has its own $\mathsf{Rx}$ value, $\mathsf{Rx_{max}}$ denotes the highest of all prescription doses in the plan, and $\mathsf{Rx_{min}}$ denotes the lowest. 
For plans with only one level of prescription dose, the above three values are equal. 
The ``no penalty'' interval for Rings and OARs (except lenses) are set to $[0, 0.5]$ and $\Bigl[0, 10/\mathsf{Rx_{min}}\Bigr]$, respectively. 
Note that Rings are normalized relative to the prescription of the target volume used to generate it, 
and the OARs are normalized using $\mathsf{Rx_{min}}$.  
The 10Gy value is chosen in the current study for all OARs because voxel doses less than 10Gy are generally insignificant for most OARs.
Further stratification of dose levels for each OAR is straightforward with our study framework. 
As an example, here we have set the ``no penalty'' interval for the lenses to be $[0, 0]$ in order to minimize the non-stochastic radiation effect. 
The actual dose distribution of a planning structure could include voxels receiving doses outside of the range $\bigl[l\cdot\mathsf{Rx}, h\cdot\mathsf{Rx}\bigr]$, 
a penalty would consequently be incurred. 
The voxels for each planning structure are binned by normalized dose $d/\mathsf{Rx}$ with a bin size of 0.002, and the relative voxel number $n/N$ is used for display in Figure~\ref{fig:reward},
where $N$ is the total number of voxels within the planning structure.  
Given the dose $d_i$ of the $i$-th bin and the number of voxels $n_i$ in this bin, 
the penalty is calculated for each planning structure as,
\begin{equation}
penalty = \sum_{i:\,\frac{d_i}{\mathsf{Rx}} < l} e^{\phi \left(l-\frac{d_i}{\mathsf{Rx}}\right)} \frac{n_i}{N}\left(l-\frac{d_i}{\mathsf{Rx}}\right) 
             + \sum_{j:\,\frac{d_j}{\mathsf{Rx}} > h} e^{\phi \left(\frac{d_j}{\mathsf{Rx}}-h\right)} \frac{n_j}{N}\left(\frac{d_j}{\mathsf{Rx}}-h\right) 
          = -score 
\label{eq:eq14}
\end{equation} 
where the first sum $\sum_i$ accumulates penalty from bins with dose less than $l\cdot\mathsf{Rx}$ and the second sum $\sum_j$ from bins with dose more than $h\cdot\mathsf{Rx}$.
The function $\phi(x) = cx$ is used to control the severity of the penalty and in our experiment $c$ is set to 20, 10, 0.5, 2, 1 for CTVs, PTVs, Rings, level-1 OARs and level-3 OARs, respectively. 
The score of a planning structure is defined as the negative of its penalty.  
Note that OARs and Rings only have the second sum, 
and $\mathsf{Rx_{min}}$ is used in place of $\mathsf{Rx}$ for OARs and Rings in plans with multiple levels of prescriptions. 
When there is no penalty, the planning structure will receive its highest score of 0. 
We also define the uniformity ($uni\!f\!orm$) for target structures as the relative dose difference between its $D_{2\%}$ and $D_{98\%}$, i.e.
\begin{equation*}
uni\!f\!orm = -\frac{D_{2\%}-D_{98\%}}{\mathsf{Rx}}  
\end{equation*}
where the uniformity is at its largest value of 0 when dose distribution within the target is completely uniform. 
We can now define the plan's score by considering all of its planning structures,
\begin{equation}
\begin{split}
score_{\mathrm{plan}} &= \sum_{i\in \mathcal{S}_1} score^i + \frac{1}{|\mathcal{S}_2|}\sum_{j \in \mathcal{S}_2} score^j  + \frac{1}{|\mathcal{S}_3|}\sum_{k \in \mathcal{S}_3} uni\!f\!orm^k
\end{split}
\label{eq:eq14_1}
\end{equation}
and calculate the reward for action $a_t$ at time step $t$ as
\begin{equation}
reward(t) = score_{\mathrm{plan}}(t+1) - score_{\mathrm{plan}}(t) 
\label{eq:eq15}
\end{equation} 
%
where $\mathcal{S}_1$ is the set of CTVs, PTVs, Rings and level-1 OARs, $\mathcal{S}_2$ is the set of level-3 OARs, and $\mathcal{S}_3$ is the set of CTVs. 

\section{RESULTS}
\subsection{Implementation details}

\begin{figure}[h]
   \begin{center}
   \includegraphics[width=0.9\textwidth]{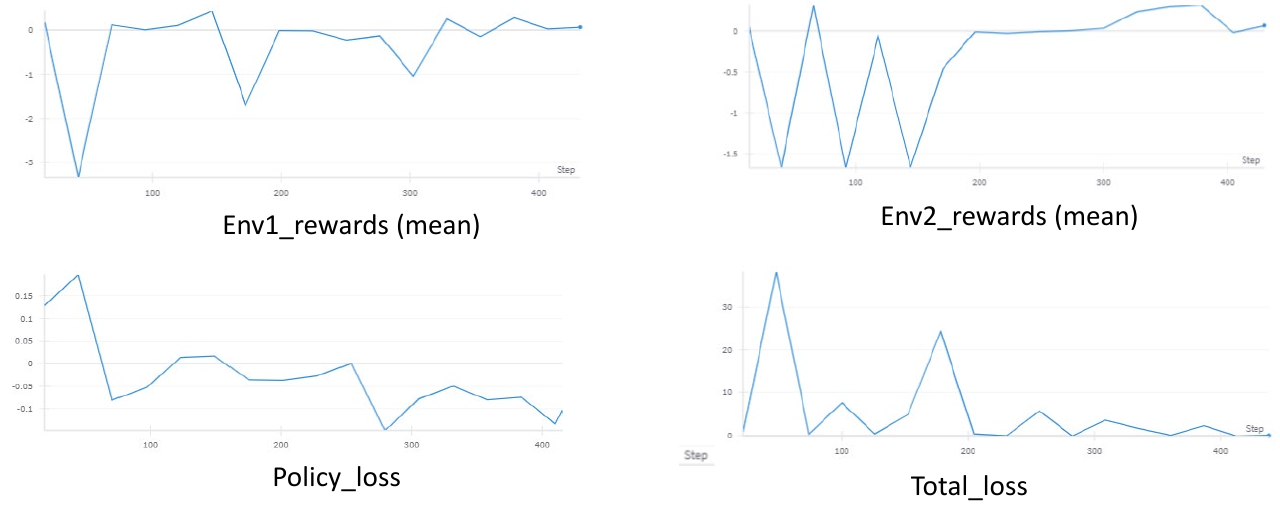}
   \caption{Rewards and losses obtained during the training procedure.    
   \label{fig:rewardloss} 
    }  
    \end{center}
\end{figure}

Our Transformer encoder consists of 4 Transformer layers, 
with hidden size, dimensions of FFN and number of heads set to 512, 1024 and 8, respectively. 
The coefficients $\alpha$, $\beta$ and the clip range $\mu$ in Eq.~\ref{eq:eq13} are set to 0.8, 0.005, and 0.3, respectively. 
An Adam optimizer~\cite{Kingma2017} with a learning rate of $5\cdot10^{-5}$ and a batch size of 12 is used to train our network.
The network is trained for 2 epochs over 30 patients and planning objectives are adjusted 4 times for each patient. 
A workstation with 2 NVIDIA RTX6000 Ada GPUs is used and the training time is about 300 hours. 
Figure~\ref{fig:rewardloss} displays the reward and loss values obtained during training. 
To improve sampling efficiency, we create two separated copies of our environment to collect samples simultaneously. 
At each step, samples collected by the two environments are gathered into one mini-batch to update the network. 
As seen clearly in Figure~\ref{fig:rewardloss}, 
the rewards are increasing and the losses are decreasing as the training progresses, 
indicating that the policy network is learning to adjust the planning objectives effectively.

\subsection{Evaluations}
\subsubsection{Trade-off behavior analysis of policy network}
\begin{figure}[t]
   \begin{center}
   \includegraphics[width=\textwidth]{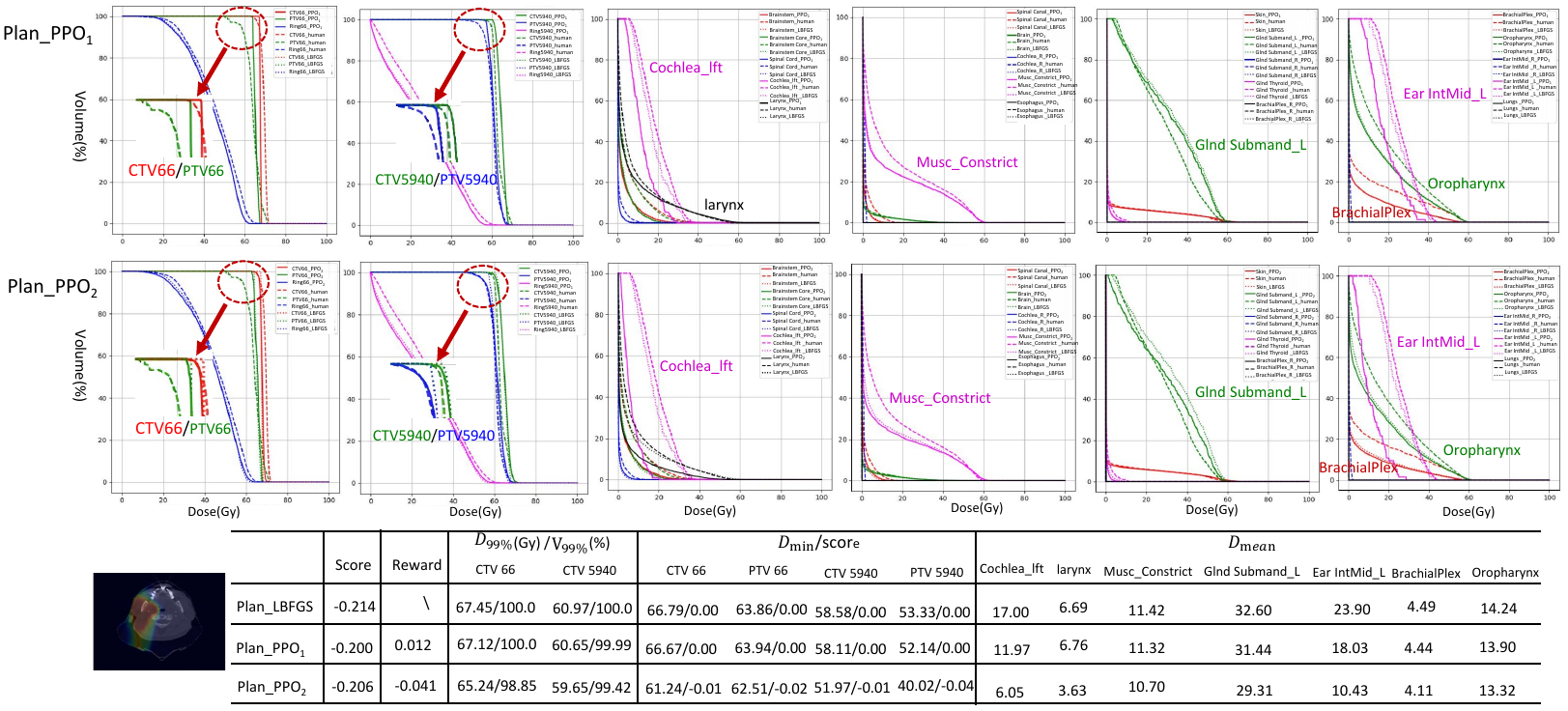}
   \caption{Plan comparison for a patient with H\&N cancer. Dash lines: Plan\_human made by human planner; Dot lines: Plan\_LBFGS generated by L-BFGS with initial planning objectives; Solid lines: Plan\_PPO$_1$ and Plan\_PPO$_2$ generated by L-BFGS with planning objectives adjusted by PPO's policy network.  
   \label{fig:behavior} 
    }  
    \end{center}
\end{figure}

To demonstrate how this policy network negotiates trade-offs during planning, 
we examine three plans: Plan\_LBFGS, Plan\_PPO$_1$ and Plan\_PPO$_2$, 
generated for a patient at three consecutive time steps. 
Plan\_LBFGS is generated by L-BFGS with initial planning objective parameters given by the empirical rules in Section~\ref{sec:EmpiricalRules}. 
Plan\_PPO$_1$ and Plan\_PPO$_2$ are generated based on Plan\_LBFGS with planning objective parameters adjusted by the PPO policy network.
By comparing Plan\_LBFGS with Plan\_PPO$_1$ and Plan\_PPO$_2$, 
we see in Figure~\ref{fig:behavior} the competing objectives of targets and OARs negotiating the resultant dose distributions. 

Emulating typical human planners' strategies,   
planning objectives are initialized with priorities given to the target volumes, i.e.
larger weights are assigned to target volumes' prescribed dose limits to begin the planning process. 
As expected, Plan\_LBFGS has the best target coverages with the highest $D_{99\%}$, $V_{99\%}$ and $D_{min}$ and relatively relaxed OAR doses. 
In subsequent steps, the PPO policy network fine-tunes planning objective parameters to enhance normal tissue sparing while minimizing the impact on target coverages.
Indeed lower OAR doses with essentially unaffected or slightly compromised CTV and PTV coverages are observed in Plan\_PPO$_1$ and Plan\_PPO$_2$, 
demonstrating the critical characteristics desired of the PPO policy network, i.e. its ability to finesse such competing priorities to within a proper range. 
This ability of our PPO network is achieved by proper designs of the rewards and the planning structures' ``no penalty'' intervals. 
It is also observed that Plan\_PPO$_2$ obtains a lower score than Plan\_PPO$_1$ because of its lower target coverages, despite significantly reduced OAR doses.
Compared with Plan\_LBFGS, 
the $D_{99\%}$ of ``CTV 66'' ($\mathsf{Rx}$=66Gy) drops from 67.5Gy to 67.1Gy and 65.2Gy in Plan\_PPO$_1$ and Plan\_PPO$_2$, respectively. 
The associated $V_{99\%}$ and $D_{min}$ remain almost unchanged in Plan\_PPO$_1$, 
but drops noticeably from 100\% to 98.9\% and from 66.8Gy to 61.2Gy in Plan\_PPO$_2$. 
The same trend also exhibits with ``CTV 5940'' ($\mathsf{Rx}$=59.4Gy). 
Therefore, even though OAR doses are lower in Plan\_PPO$_2$ than in Plan\_PPO$_1$, 
Plan\_PPO$_2$ still receives a slightly negative reward of -0.041 compared with Plan\_PPO$_1$ due to its compromised target coverages. 
This is aligned with human intuition and planning preferences. 
Essentially, the policy network is searching for the optimal planning objective parameters by learning the adjustment intelligence from the training process.

\subsubsection{Comparison with plans made by human planners}

\begin{figure}[htbp]
    \centering
    \begin{subfigure}[b]{\textwidth}
        \centering
        \includegraphics[width=0.97\textwidth]{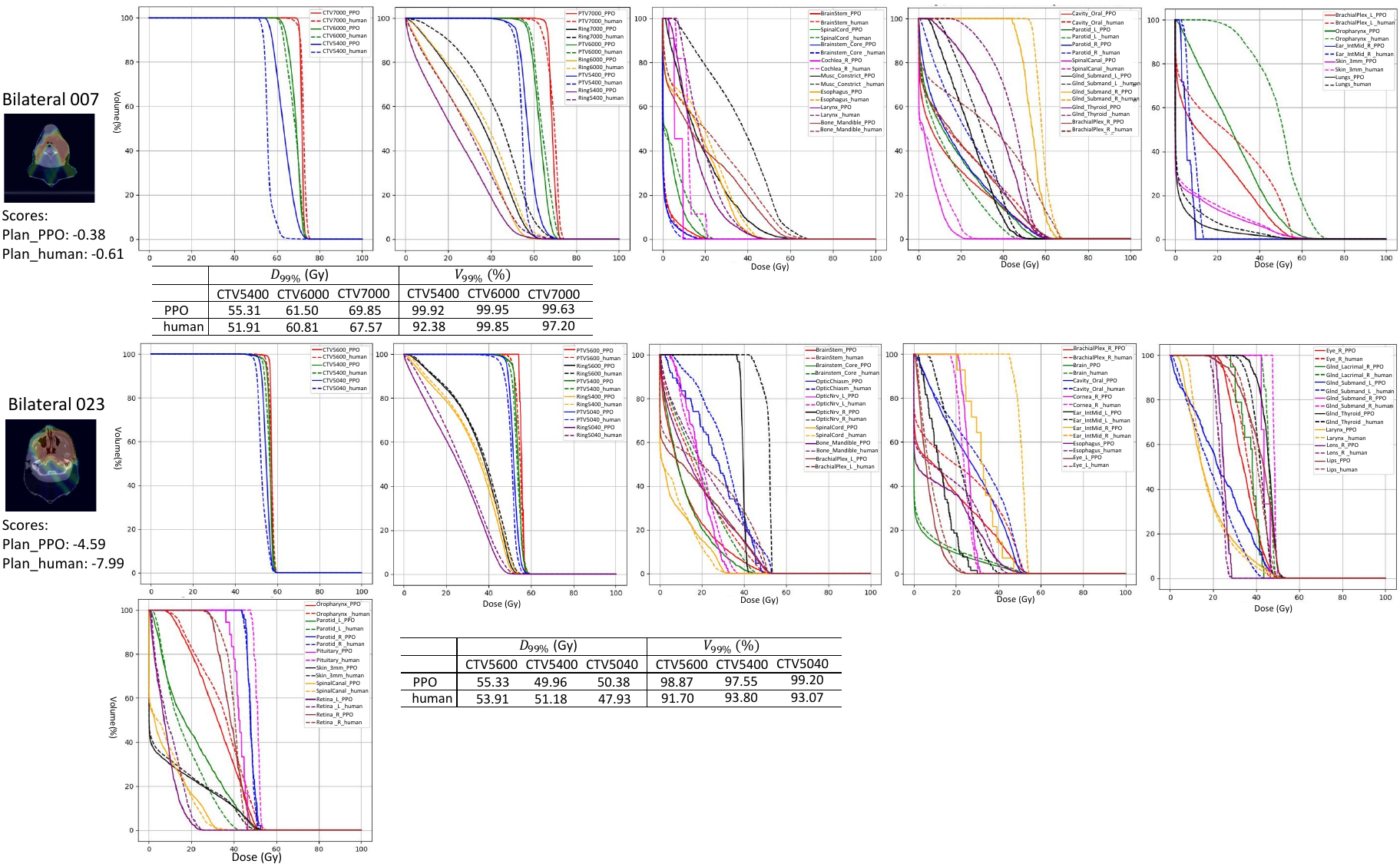}
        \caption{Patients with bilateral H\&N cancer}
        \label{fig:first}
    \end{subfigure}
		\begin{subfigure}[b]{\textwidth}
        \centering
        \includegraphics[width=0.97\textwidth]{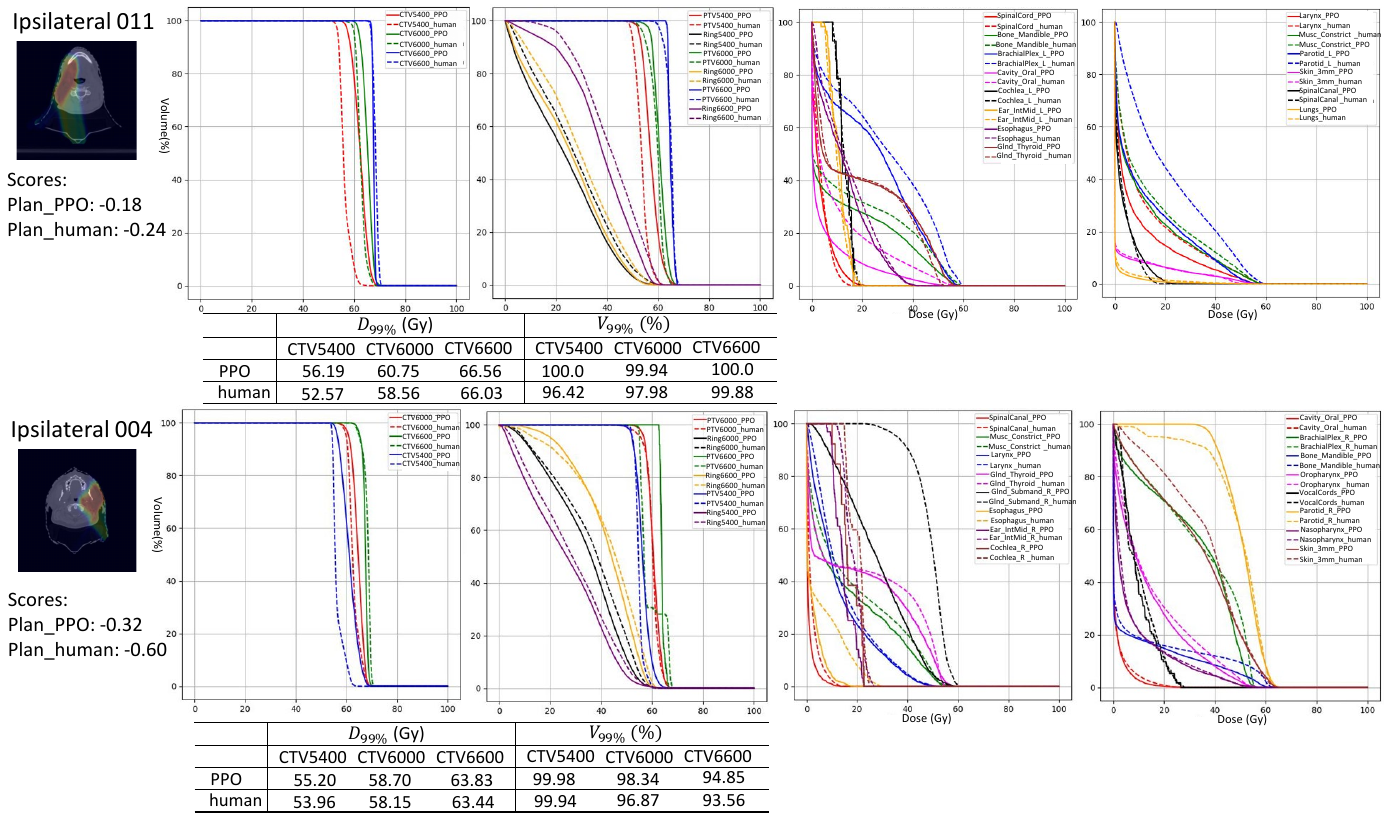}
        \caption{Patients with ipsilateral H\&N cancer}
        \label{fig:second}
    \end{subfigure}
		\caption{Comparison with human plans on test patients with H\&N cancer. Solid lines: Plan\_PPO yielded by proposed method; Dash lines: Plan\_human made by human planners. For each patient, the first DVH figure is for CTVs, the second one is for PTVs and Rings, and the others are for OARs.}
    \label{fig:BilIpsi}
\end{figure}

Human planners employ a wide variety of techniques and strategies when optimizing treatment plans. 
Different margin sizes, objective weights and dose limits are often used by individual planners.  
Large variability therefore exists amongst treatment plans optimized by human planners. 
By contrast, our model uses a consistent strategy for all test patients. 
Resultant plans are compared with those made by human planners in Figure~\ref{fig:BilIpsi}.

Using the definition of plan scores given in Eq.~\ref{eq:eq14_1}, 
Plan\_PPO achieves scores of -0.38, -4.59, -0.18 and -0.32 on the four test patients, 
compared with the scores of -0.61, -7.99, -0.24 and -0.60 for Plan\_human. 
Apparently, plans yielded by our model surpass human plans with higher scores for both bilateral and ipsilateral H\&N cancers.
Using $D_{99\%}$ and $V_{99\%}$ as the criteria for target coverages,
Plan\_PPO also outperforms Plan\_human on all test patients. 
Specifically on the three target volumes: CTV5400, CTV6000 and CTV7000 of patient ``Bilateral 007", 
Plan\_PPO achieves higher $D_{99\%}$ of 55.31Gy, 61.50Gy and 69.85Gy, and higher $V_{99\%}$ of 99.92\%, 99.95\% and 99.63\% than Plan\_human, 
whose $D_{99\%}$ and $V_{99\%}$ are 51.91Gy, 60.81Gy and 67.57Gy, and 92.38\%, 99.85\% and 97.20\%, respectively. 
For the more challenging case ``Bilateral 023", 
human planners have to make significant compromise on target coverages to protect multiple OARs from being radiated at high doses. 
Therefore, Plan\_human obtains relatively low $D_{99\%}$ of 53.91Gy, 51.18Gy and 47.93Gy, 
and relatively low $V_{99\%}$ of 91.70\%, 93.80\% and 93.07\% on the target volumes CTV5600, CTV5400 and CTV5040, respectively. 
By contrast, the $D_{99\%}$ and $V_{99\%}$ are 55.33Gy, 49.96Gy and 50.38Gy, and 98.87\%, 97.55\% and 99.20\% for Plan\_PPO.
Though Plan\_PPO has a lower $D_{99\%}$ on CTV5040 (49.96Gy vs. 51.18Gy), 
it outperforms Plan\_human on other two target volumes (55.33Gy vs. 53.91Gy and 50.38Gy vs. 47.93Gy), 
and achieves much better $V_{99\%}$ on all three target volumes (98.87\% vs. 91.70\%, 97.55\% vs. 93.80\%, 99.20\% vs. 93.07\%). 
Overall, Plan\_PPO achieves better target coverages than Plan\_human on this difficult bilateral H\&N case.  
For patient ``Ipsilateral 011", 
the $D_{99\%}$ and $V_{99\%}$ of Plan\_PPO are 56.19Gy, 60.75Gy and 66.56Gy, and 100.0\%, 99.94\% and 100.0\% on the three target volumes, respectively; compared with Plan\_human's $D_{99\%}$ of 52.57Gy, 58.56Gy and 66.03Gy, and $V_{99\%}$ of 96.42\%, 97.98\% and 99.88\%. 
Given the prescription doses of 54Gy, 60Gy and 66Gy, Plan\_PPO satisfies our clinical requirements $D_{99\%}>99\%\cdot\mathsf{Rx}$ and $V_{99\%}>99\%$ on all target volumes, but Plan\_human fails on two of them. 
For the last patient ``Ipsilateral 004", Plan\_PPO and Plan\_human obtain similar $D_{99\%}$, but Plan\_PPO achieves higher $V_{99\%}$ on CTV6000 and CTV6600, i.e., 98.34\% and 94.85\%; compared with the 96.87\% and 93.56\% of Plan\_human. 
Therefore, in terms of target coverages, Plan\_PPO performed superiorly than Plan\_human on all 4 test patients.  

In addition to superior target coverages, 
Plan\_PPO also surpasses Plan\_human in terms of better OAR sparing on all 4 test cases. 
Specifically, patient ``Bilateral 007" has a total of 21 OARs, 
of which 18 receive significantly lower doses in Plan\_PPO than in Plan\_human, including ``SpinalCord", ``Musc\_Constrict", ``Larynx", ``Glnd\_Submand", ``Glnd\_Thyroid", ``BrachialPlex", ``Oropharynx", etc. 
Although ``BrainStem", ``Brainstem\_Core" and ``Parotid\_L" receive higher doses in Plan\_PPO, the maximal doses of ``BrainStem" and ``Brainstem\_Core" are lower than 20Gy and the mean dose of ``Parotid\_L" is only 16.82Gy, which are all clinically acceptable. 
For patient ``Ipsilateral 011", ``Bone\_Mandible", ``BrachialPlex\_L", ``Cavity\_Oral", ``Esophagus", ``Glnd\_Thyroid", ``Larynx", ``Musc\_Constrict" and ``Parotid\_L" receive pertinent doses.
Seven of these 8 OARs receive significantly lower doses in Plan\_PPO than in Plan\_human except ``Glnd\_Thyroid", for which the mean dose is 18.98Gy in Plan\_PPO and 18.72Gy in Plan\_human. 
Both values are well below thyroid gland's clinical requirement of $D_{mean}<45$Gy. 
Therefore, considering the overall target coverages and OAR sparing, 
it is determined that Plan\_PPO outperforms Plan\_human on patient ``Ipsilateral 011".
For patient ``Ipsilateral 004", 
Plan\_PPO achieves lower doses on 15 OARs and Plan\_human on one OAR, ``Parotid\_R". 
Closer examination reveals that ``Glnd\_Submand\_R" and ``Parotid\_R" have similar mean dose values in Plan\_human at 48.93Gy and 49.03Gy, respectively. 
Plan\_PPO suppresses the mean dose of ``Glnd\_Submand\_R" to 27.39Gy at a cost of slightly increased ``Parotid\_R" mean dose of 50.98Gy. 
Given the clinical requirements $D_{mean}<35$Gy for ``Glnd\_Submand'' and $D_{mean}<26$Gy for ``Parotid'', 
overall Plan\_PPO achieves superior performance than Plan\_human.  
For the most challenging case ``Bilateral 023", about 30 OARs receive pertinent doses. 
Amongst those received significant doses, ``OpticNrv\_R", ``OpticChiasm", ``Ear\_IntMid\_R", ``Glnd\_Lacrimal\_R", ``Glnd\_Submand\_R" and ``Pituitary" receive the maximal doses of 53.52Gy, 53.03Gy, 54.16Gy, 49.70Gy, 49.82Gy and 53.44Gy in Plan\_human, 
with associated mean doses of 51.25Gy, 31.58Gy, 50.79Gy, 46.00Gy, 48.49Gy and 51.01Gy, respectively. 
By contrast, in Plan\_PPO the maximal doses of the above OARs are significantly reduced to 41.99Gy, 48.63Gy, 47.23Gy, 43.55Gy, 46.50Gy and 46.52Gy, and the mean doses are also reduced to 39.52Gy, 27.19Gy, 32.71Gy, 36.90Gy, 44.14Gy and 42.52Gy, respectively. 
For the OARs receiving moderate doses, 
Plan\_PPO outperforms Plan\_human on ``BrainStem", ``Brainstem\_Core", ``Bone\_Mandible", ``BrachialPlex\_L", ``BrachialPlex\_R", ``Brain", ``Cavity\_Oral", ``Ear\_IntMid\_L", ``Esophagus", ``Eye\_L", ``Eye\_R", ``Retina\_L", ``Retina\_R" and ``Oropharynx", 
while Plan\_human performs better on ``Glnd\_Thyroid", ``Glnd\_Submand\_L" and ``Parotid\_L". 
For the remaining 7 OARs, both Plan\_PPO and Plan\_human achieve dose levels well below clinical requirements. 
Plan\_PPO is again preferred over Plan\_human for this challenging case. 

In summary, compared with human planners, the PPO model achieves clinically preferable trade-offs between target coverages and OAR sparing by finessing the adjustments of planning objective parameters. 
The policy network has learned to avoid aggressive adjustments, successfully lowered OAR doses and maintained target coverages well within reasonable ranges. 
Consequently, high-quality patient treatment plans can be automatically generated by the proposed method with minimum human interaction. 

\subsubsection{Model generalizability}

\begin{figure}[h]
   \begin{center}
   \includegraphics[width=0.7\textwidth]{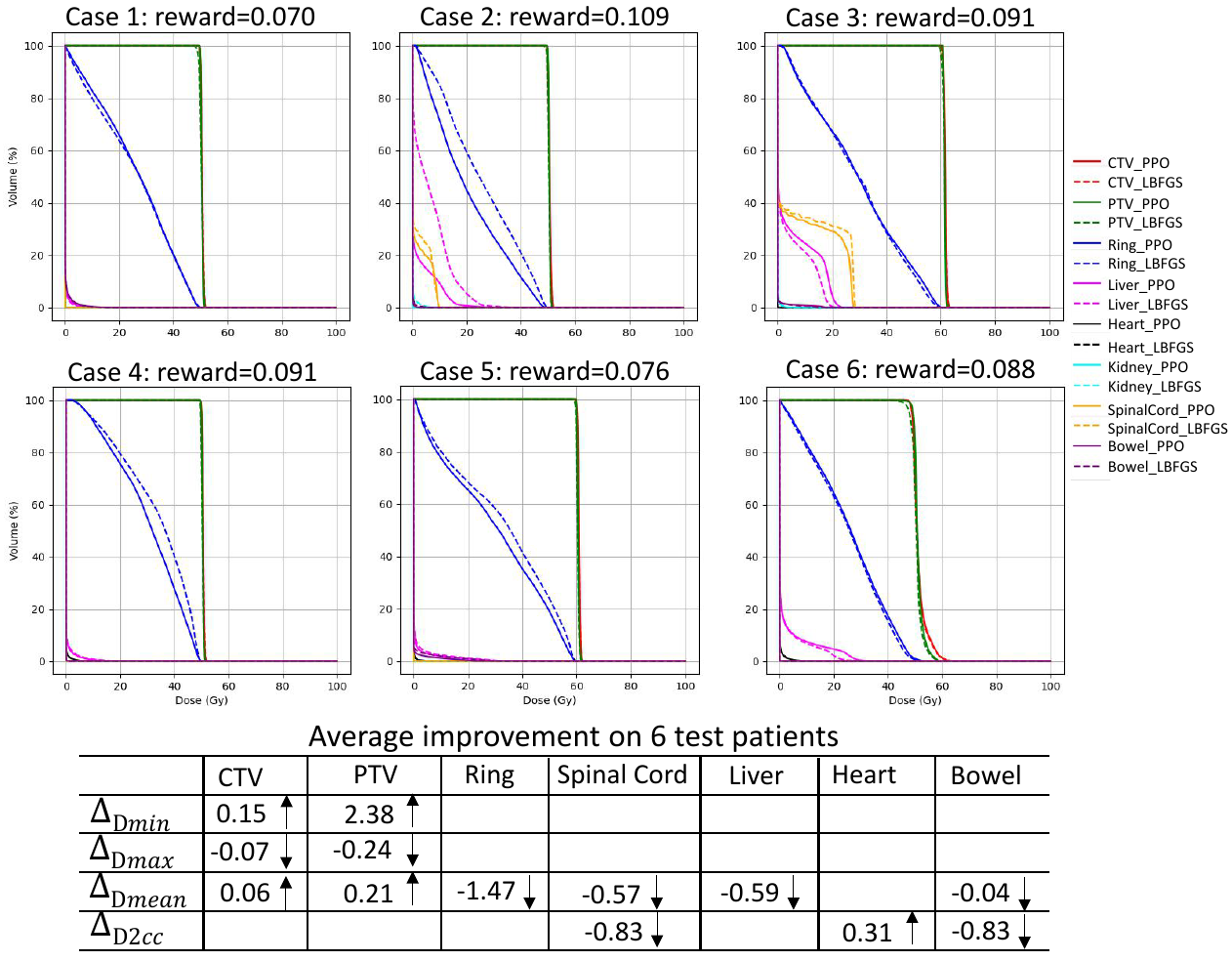}
   \caption{Results on test patients with liver cancer. Solid lines: Plan\_PPO obtained with planning objective parameters produced by PPO's policy network; Dash lines: Plan\_LBFGS obtained with initial planning objective parameters. Rewards shown here are obtained by comparing Plan\_PPO and Plan\_LBFGS. $\Delta_{Dmin}$, $\Delta_{Dmax}$, $\Delta_{Dmean}$ and $\Delta_{D2cc}$ represent average changes of $D_{min}$, $D_{max}$, $D_{mean}$ and $D_{2cc}$ on all test patients.    
   \label{fig:liver} 
    }  
    \end{center}
\end{figure}
To verify the generalizability of the proposed method, a liver model using the same algorithm and adapted parameters is implemented. 
For liver plans, CTV to PTV expansion is set to 5mm and a 15mm expansions from the Gap is used to create the Ring.
The OARs considered in this liver model only include ``Spinal Cord'', ``Liver'', ``Heart'', ``Kidneys'' and ``Bowel''. 
Since liver plans involve less OARs than H\&N, 
robust optimization using L\_BFGS becomes computationally feasible due to more lenient memory requirements~\cite{A30}.  
A total of 21 robust scenarios are considered with a setup uncertainty of 5mm and a range uncertainty of 3.5\%.
The worst case from the 21 robust scenarios is used to calculate the loss in Eq.~\ref{eq:eq1} for planning structures.

As shown in Figure~\ref{fig:liver}, 
all six patients achieve positive rewards after adjusting planning objective parameters using PPO's policy network, 
indicating the effectiveness of the policy network. 
On average, the $D_{min}$ and $D_{mean}$ of CTVs are increased by 0.15Gy and 0.06Gy, respectively, 
and the $D_{max}$ of CTVs is decreased by 0.07Gy. 
Similarly, the $D_{min}$ and $D_{mean}$ of PTVs are increased by 2.38Gy and 0.21Gy, respectively, 
and the $D_{max}$ of PTVs is decreased by 0.24Gy.
In addition, the $D_{mean}$ of Rings and ``Liver'' are significantly reduced by 1.47Gy and 0.59Gy, respectively, 
and the $D_{2cc}$ of ``Spinal Cord'' and ``Bowel'' are both decreased by 0.83Gy. 
Only one of the OAR doses, the averaged $D_{2cc}$ of ``Heart'', has seen a minor increase of 0.31Gy.
This is likely because the values of ``Heart'' $D_{mean}$ are always below 10Gy, well within the allowed OAR dose interval designed in the model, and the policy network therefore sees no need to push heart dose lower in these cases. 
In summary, the adjustments of planning objective parameters made by PPO's policy network can indeed improve OAR dose statistics without sacrificing target coverages in proton PBS liver planning. 
The proposed treatment planning framework that automates complex H\&N PBS treatment planning can indeed be generalized to other simpler sites such as liver.  

\section{DISCUSSION}

Recently, DRL-based treatment planning has been proposed to alleviate human planners' workload and improve plan quality. 
However, prior studies are developed based upon Q-learning~\cite{A13}, 
which can only predict actions in low-dimensional discrete spaces. 
As a result, they suffer from poor scalability and flexibility, 
and cannot be used for complex systems such as H\&N treatment planning where a large number of target volumes and OARs are involved. 
In this study, a PPO-based treatment planning framework automatically iterates the planning objective parameters in a continuous action space using a policy network for proton H\&N treatment plans. 
The PPO network is trained under the guidance of a novel dose distribution-based reward function. 
Proposed method achieves human-level performance without human interference and is successfully adapted to other cancer types.
     
This pilot study has the following limitations. 
Firstly, L-BFGS is specifically designed for optimization problems involving large numbers of variables but limited by available memories. 
Therefore, it often has a prolonged computation time. 
More efficient optimization engine should be investigated for future works. 
The second limitation in the current study is that all OARs except ``Lens'' are scored with the same ``no penalty'' interval of $[0, 10/\mathsf{Rx}]$ and penalized with the same severity function $\phi$ in Eq.~\ref{eq:eq14}. 
However, individual OARs have different dose tolerances. 
Stratifying training parameters for individual OARs will further enhance this framework by aligning the reward function more closely with clinical criteria.
Despite these limitations, the proposed method is a pioneering step toward automatic treatment planning as it achieves promising human-level performance for complex treatment sites. 

\section{CONCLUSION}
To the best of our knowledge this is the first study using policy gradient in DRL to automate radiotherapy treatment planning. 
An advanced policy gradient algorithm, PPO, is used to train a Transformer-based policy network for planning objective adjustments. 
This model overcomes the scalability and flexibility limitations seen in prior Q-learning-based studies by enabling continuous adjustments on a large number of planning objectives. 
It can therefore be used for complex treatment sites. 
Treatment plans generated by the proposed model demonstrate dose distributions comparable or superior to those produced by human planners for both bilateral and ipsilateral H\&N cancers.
Moreover, the proposed method is also successfully generalized to liver cancer. 
This study has demonstrated the potential of using DRL to fully automate proton PBS treatment planning.   

\section{CONFLICT OF INTEREST STATEMENT}
This study is funded by a grant from Varian Medical Systems.

\section*{REFERENCES}
\addcontentsline{toc}{section}{\numberline{}References}
\vspace*{-20mm}




\begin{thebibliography}{10}
\bibitem{A04}
R.~Lu, R.~J. Radke, L.~Happersett, J.~Yang, C.-S. Chui, E.~Yorke, and
  A.~Jackson,
\newblock {Reduced-order parameter optimization for simplifying prostate IMRT
  planning},
\newblock Phys. Med. Biol. {\bf 52}, 849 -- 870 (2007).

\bibitem{A05}
H.~Wang, P.~Dong, H.~Liu, and L.~Xing,
\newblock {Development of an autonomous treatment planning strategy for
  radiation therapy with effective use of population-based prior data},
\newblock Med. Phys. {\bf 44}, 389 -- 396 (2017).

\bibitem{A06}
H.~Yan, F.~Yin, H.~Guan, and J.~H. Kim,
\newblock {AI-guided parameter optimization in inverse treatment planning},
\newblock Phys. Med. Biol. {\bf 48}, 3565 -- 3580 (2003).

\bibitem{A07}
T.~Lee, M.~Hammad, T.~C.~Y. Chan, T.~Craig, and M.~B. Sharpe,
\newblock {Predicting objective function weights from patient anatomy in
  prostate IMRT treatment planning},
\newblock Med. Phys. {\bf 40}, 121706--3 (2013).

\bibitem{A08}
C.~Shen, Y.~Gonzalez, P.~Klages, N.~Qin, H.~Jung, L.~Chen, D.~Nguyen, S.~B.
  Jiang, and X.~Jia,
\newblock {Intelligent inverse treatment planning via deep reinforcement
  learning, a proof-of-principle study in high dose-rate brachytherapy for
  cervical cancer},
\newblock Phys. Med. Biol. {\bf 64}, 115013 (2019).

\bibitem{A09}
C.~Shen, D.~Nguyen, L.~Chen, Y.~Gonzalez, R.~McBeth, N.~Qin, S.~B. Jiang, and
  X.~Jia,
\newblock {Operating a treatment planning system using a deep-reinforcement
  learning-based virtual treatment planner for prostate cancer
  intensity-modulated radiation therapy treatment planning},
\newblock Med. Phys. {\bf 47}, 2329 -- 2336 (2020).

\bibitem{A10}
C.~Shen, L.~Chen, and X.~Jia,
\newblock {A hierarchical deep reinforcement learning framework for intelligent
  automatic treatment planning of prostate cancer intensity modulated radiation
  therapy},
\newblock Phys. Med. Biol. {\bf 66}, 134002 (2021).

\bibitem{A11}
G.~Pu, S.~Jiang, Z.~Yang, Y.~Hu, and Z.~Liu,
\newblock {Deep reinforcement learning for treatment planning in high-dose-rate
  cervical brachytherapy},
\newblock Physica Medica {\bf 94}, 1 -- 7 (2022).

\bibitem{A16}
H.~Wang, X.~Bai, Y.~Wang, Y.~Lu, and B.~Wang,
\newblock {An integrated solution of deep reinforcement learning for automatic
  IMRT treatment planning in non-small-cell lung cancer},
\newblock Front. Oncol. {\bf 13}, 1124458 (2023).

\bibitem{A18}
W.~Hrinivich and J.~Lee,
\newblock {Artificial intelligence-based radiotherapy machine parameter
  optimization using reinforcement learning},
\newblock Med. Phys. {\bf 47}, 6140--50 (2020).

\bibitem{A19}
J.~Zhang, C.~Wang, Y.~Sheng, M.~Palta, B.~Czito, C.~Willett, J.~Zhang, P.~J.
  Jensen, F.-F. Yin, Q.~Wu, Y.~Ge, and Q.~J. Wu,
\newblock {An interpretable planning bot for pancreas stereotactic body
  radiation therapy},
\newblock Int. J. Radiat. Oncol. Biol. Phys. {\bf 109}, 1076 -- 1085 (2021).

\bibitem{A13}
C.~J. C.~H. Watkins and P.~Dayan,
\newblock {Q-learning},
\newblock Machine Learning {\bf 8}, 279 -- 292 (1992).

\bibitem{A22}
J.~Schulman, F.~Wolski, P.~Dhariwal, A.~Radford, and O.~Klimov,
\newblock {Proximal policy optimization algorithms},
\newblock arXiv preprint , arXiv:1707.06347 (2017).

\bibitem{A24}
A.~Vaswani, N.~Shazeer, N.~Parmar, J.~Uszkoreit, L.~Jones, A.~N. Gomez,
  L.~Kaiser, and I.~Polosukhin,
\newblock {Attention is all you need},
\newblock Advances in Neural Info. Proc. Syst. {\bf 30} (2017).

\bibitem{Frank2014}
S.~J. Frank et~al.,
\newblock Multifield optimization intensity modulated proton therapy for head
  and neck tumors: a translation to practice,
\newblock Int. J. Radiat. Oncol. Biol. Phys. {\bf 89}, 846--853 (2014).

\bibitem{A21}
K.~Leach, S.~Tang, J.~Sturgeon, A.~K. Lee, R.~Grover, P.~Sanghvi, J.~Urbanic,
  and C.~Chang,
\newblock {Beam-specific spot guidance and optimization for PBS proton
  treatment of bilateral head and neck cancers},
\newblock Int. J. Particle Therapy {\bf 8}, 50 -- 61 (2021).

\bibitem{A29}
A.~Hatamizadeh, Y.~Tang, V.~Nath, D.~Yang, A.~Myronenko, B.~Landman, H.~R.
  Roth, and D.~Xu,
\newblock {UNETR: Transformers for 3D medical image segmentation},
\newblock IEEE Winter Conf. on Appl. of Comp. Vision , 574 -- 584 (2022).

\bibitem{Unkelbach2016}
J.~Unkelbach, P.~Botas, D.~Giantsoudi, B.~L. Gorissen, and H.~Paganetti,
\newblock {Reoptimization of intensity modulated proton therapy plans based on
  linear energy transfer},
\newblock Int. J. Radiat. Oncol. Biol. Phys. {\bf 96}, 1097--1106 (2016).

\bibitem{Ma2018}
J.~Ma, H.~S. Wan Chan~Tseung, M.~G. Herman, and C.~Beltran,
\newblock {A robust intensity modulated proton therapy optimizer based on Monte
  Carlo dose calculation},
\newblock Med. Phys. {\bf 45}, 4045--4054 (2018).

\bibitem{A23}
D.~C. Liu and J.~Nocedal,
\newblock {On the limited memory BFGS method for large scale optimization},
\newblock Math. Programming {\bf 45}, 503 -- 528 (1989).

\bibitem{A33}
T.~Haarnoja, A.~Zhou, P.~Abbeel, and S.~Levine,
\newblock {Soft actor-critic: off-policy maximum entropy deep reinforcement
  learning with a stochastic actor},
\newblock Int. Conf. on Machine Learning , 1861 -- 1870 (2018).

\bibitem{A26}
J.~Schulman, P.~Moritz, S.~Levine, M.~I. Jordan, and P.~Abbeel,
\newblock {High-dimensional continuous control using generalized advanced
  estimation},
\newblock arXiv preprint , arXiv:1506.02438 (2018).

\bibitem{Kingma2017}
D.~P. Kingma and J.~Ba,
\newblock {Adam: a method for stochastic optimization},
\newblock arXiv preprint , arXiv:1412.6980 (2017).

\bibitem{A30}
A.~Fredirksson, A.~Forsgren, and B.~Hoardemark,
\newblock {Minimax optimization for handling range and setup uncertainties in
  proton therapy},
\newblock Med. Phys. {\bf 38}, 1672 -- 1684 (2011).

\end{thebibliography}



\bibliographystyle{./medphy.bst}    


\end{document}